\documentclass[aps,prb,twocolumn,superscriptaddress]{revtex4}

\usepackage{graphicx, color}% Include figure files

\begin{document}

%\preprint{submitted to PRB}

\title{Mechanical properties of {Pt} monatomic chains}
%\title{Mechanical properties of atomic-sized platinum contact}

\author{T. Shiota}

%\thanks{}
\altaffiliation{Present address: Department of Metallurgy and Ceramics Science, Graduate school of Science and Engineering, Tokyo Institute of Technology, 2-12-1-S7-14 Ookayama, Meguro-ku, Tokyo 152-8552, Japan. E-mail address: tshiota@ceram.titech.ac.jp}
\affiliation{Kamerlingh Onnes Laboratorium, Universiteit Leiden, Postbus 9504, 2300 RA Leiden, The Netherlands}

\author{A. I. Mares}
%\thanks{}
\affiliation{Kamerlingh Onnes Laboratorium, Universiteit Leiden, Postbus 9504, 2300 RA Leiden, The Netherlands}

\author{A. M. C. Valkering}
%\thanks{}
\affiliation{Kamerlingh Onnes Laboratorium, Universiteit Leiden, Postbus 9504, 2300 RA Leiden, The Netherlands}

\author{T. H. Oosterkamp}
%\thanks{}
\affiliation{Kamerlingh Onnes Laboratorium, Universiteit Leiden, Postbus 9504, 2300 RA Leiden, The Netherlands}

\author{J. M. van Ruitenbeek}
%\thanks{}
\altaffiliation{E-mail address: ruitenbeek@physics.leidenuniv.nl}
\affiliation{Kamerlingh Onnes Laboratorium, Universiteit Leiden, Postbus 9504, 2300 RA Leiden, The Netherlands}

%\date{\today}

\begin{abstract}
The mechanical properties of platinum monatomic chains were
investigated by simultaneous measurement of an effective stiffness
and the conductance using our newly developed mechanically
controllable break junction (MCBJ) technique with a tuning fork as
a force sensor. When stretching a monatomic contact (two-atom
chain), the stiffness and conductance increases at the early stage
of stretching and then decreases just before breaking, which is
attributed to a transition of the chain configuration and bond
weakening. A statistical analysis was made to investigate the
mechanical properties of monatomic chains. The average 
stiffness shows minima at the peak positions of the
length-histogram. From this result we conclude that the peaks in
the length-histogram are a measure of the number of atoms in the
chains, and that the chains break from a strained state.
Additionally, we find that the smaller the initial stiffness of
the chain is, the longer the chain becomes. This shows that softer
chains can be stretched longer.
\end{abstract}

%\pacs{}

%\keywords{}

\maketitle

\section{Introduction}
Since the formation of Au monatomic chains was first demonstrated
in 1998 \cite{yanson98, ohnishi98}, the chain formation in other
materials and their quantum properties have attracted great
interest of many researchers. Until now, many studies have been
reported on this subject (for a review see
Ref.~\onlinecite{agrait03}), and especially the electrical
properties of the chains have been investigated from the
experimental and theoretical points of view. In many cases, the
chains were characterized by information coming from electrical
measurements, for example, conductance histogram, I-V
spectroscopy, shot noise, etc \cite{agrait03,agrait02,smit03}. In
contrast to the electrical properties, the mechanical properties
of the chains have not been studied much, although these two
properties are closely related \cite{stafford98, stafford99}.
Understanding their mechanical properties is essential for further
understanding of the chain formation, for the realization of
manipulation of atoms and molecules, and for elucidation of
fundamental physics of tribological phenomena such as friction,
adhesion and wear.

The mechanical properties of atomic chains have been studied
experimentally and theoretically for Au. Rubio-Bollinger and
coworkers used a combination of a scanning tunnelling microscope
(STM) with an atomic force microscope (AFM) to measure
simultaneously the conductance of the chains and the tensile force
for elongation and breaking of the chains \cite{rubio96,rubio01}.
Recently, a new experimental system has been independently
developed by our group \cite{valkering05} and by Rubio-Bollinger
\textit{et al.} \cite{rubio04} to study directly the stiffness of
the chains. Theoretical calculations and molecular dynamic
simulations
\cite{Torres99,Todorov96,sorensen98,sanchez99,hakkinen00,Silva01,rubio01,Dreher05}
have been performed to interpret the experimental results on Au
chains. However, few experimental studies have been reported on
the mechanical properties of the other chain forming metals.

Chain formation has been demonstrated experimentally \cite{smit01,
smit03} and studied theoretically \cite{bahn01,garcia05,pauly06}
for two other 5d elements, namely Pt and Ir. In this study, we
address for the first time the formation and stability of Pt
monatomic chains by measuring the mechanical stiffness of the
nanowires. Using our newly developed system, the stiffness and
conductance of Pt chains can be measured simultaneously, and the
data can be analyzed by a statistical method.

\section{Experimental technique}
In order to simultaneously measure the stiffness and conductance
of Pt monatomic chains, we used a mechanically controllable break
junction (MCBJ) technique combined with a miniature tuning fork as
a force sensor, as described in Ref.~\onlinecite{valkering05}.
Compared to the STM-AFM system which has been used in most
experiments on microscopic force measurement, the advantages of
our system are as follows: (1) our system increases mechanical
stability, permitting a statistical analysis on measurements of
many different monatomic chains; (2) the tuning fork has a much
higher stiffness (of the order of $10^4$ N/m) than an
AFM-cantilever (typically a few tens N/m). This ensures that the
formation and stability of the monatomic chains are not influenced
by the motion of the force sensor. (3) In addition, the stiffness
of the tuning fork allows the use of a dynamic force measurement
technique where the tuning fork is excited at its resonance
frequency. In AFM experiments this is only possible at amplitudes
comparable to or larger than the interatomic distance.

\begin{figure}[!b]
\includegraphics[width=7cm]{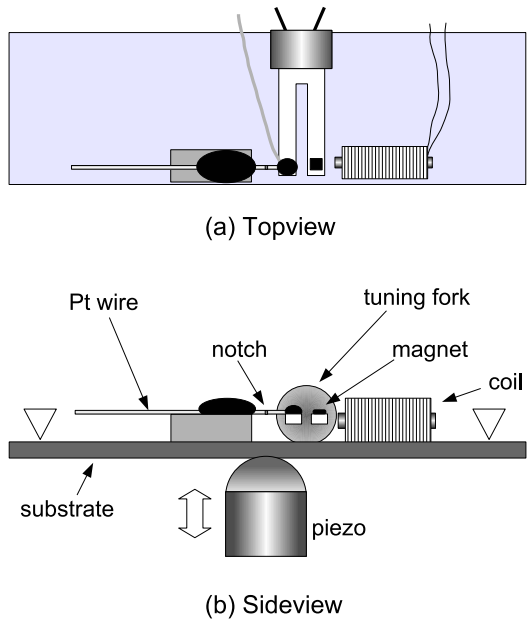}
\caption{Schematic drawings of (a) top-view and (b) side-view of
our MCBJ system with a tuning fork as a force sensor. See text for
further details.} \label{sample_config}
\end{figure}

Figure \ref{sample_config} shows schematic drawings of (a) the
top-view and (b) a side-view of the sample configuration in our
MCBJ system. We used commercial tuning forks. The base of the
tuning fork is soldered on a phosphorus-bronze substrate with both
prongs freely standing. One end of the notched Pt wire with a
diameter of 100 ${\mu}$m is fixed on a small metallic block placed
at the same height as the prongs of the tuning fork. The other end
is glued on one prong of the tuning fork. The notch of the wire is
located  between the prong and block. On the other free prong, a
magnet is glued, and a coil is fixed on the substrate next to the
tuning fork. The tuning fork can be mechanically excited by an ac
magnetic field generated by the coil.

The sample was mounted in an insert that is pumped down to a
pressure of $10^{-5}$mbar at room temperature before being cooled
down to liquid helium temperatures. The sample is elongated by
bending the substrate until it is broken. By relaxing the
substrate, the wire can be brought back into contact. By adjusting
the voltage applied to the piezo-electric element the size of the
contact can be adjusted reversibly on the atomic scale. The
notched wire is only broken once the system is under cryogenic
vacuum, ensuring clean surfaces in this way.

The tuning fork coupled with the contact is oscillated at its
resonance frequency with an amplitude lower than 6 pm (peak to
peak). Since the oscillation amplitude is kept two orders of
magnitude smaller than the inter-atomic distance for Pt the
oscillation does not influence the formation and stability of the
contact. The resonance frequency of the tuning fork coupled
through the contact will be shifted from that for a broken
contact. The frequency shift ($\Delta f$) depends on the contact
stiffness ($\Delta k$), and $\Delta f$ can be related to $\Delta
k$ using a harmonic oscillator model as:
\begin{equation}
{\Delta k} = 2 k\frac{\Delta f}{ f},
\end{equation}
where $f$ is the resonance frequency of the tuning fork for the
broken contact, and $k$ the stiffness of the free tuning fork.
$\Delta f$ was measured using two methods: (1) with a phase-locked
loop (PLL) technique and (2) by measuring directly the phase shift
($\Delta \phi$) at a constant frequency and by relating it to
$\Delta f$, assuming that the quality factor $Q$ is not changed by
breaking the contact. We have confirmed that both methods give
similar results. In most of the cases we have used the faster
second method.
\begin{figure}[!t]
\includegraphics[width=7cm]{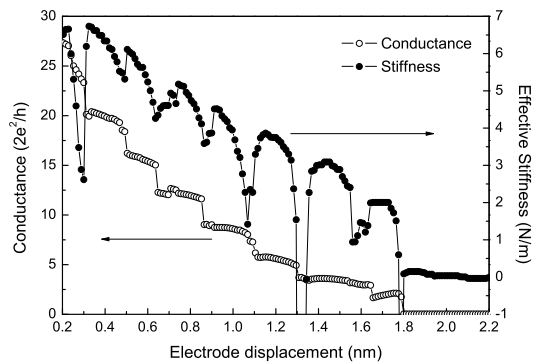}
\caption{Simultaneous measurement of the stiffness and conductance
in the process of breaking an atomic-sized Pt contact.}
\label{nanoPt}
\end{figure}
Simultaneously with $\Delta f$ we record the conductance of the
contact. A constant bias voltage of 10--50 mV is applied to the
contact and the resulting current is measured using a home-built
current-to-voltage converter. The data acquisition and the
measurement control are done by using a 16 bit analog-to-digital
converter card, using home-built LabView software \cite{LabView}.

\section{Results and Discussion}
\subsection{Nano-sized contact}
Figure \ref{nanoPt} shows typical traces of the stiffness and
conductance measured simultaneously in the process of breaking a
nano-sized Pt contact. The conductance decreases step-wise with
increasing electrode displacement, i.e., the conductance trace is
formed by plateaus separated by sudden drops. As reported
previously, the contact evolution in the breaking process is
composed of two alternating stages \cite{rubio96}: (1) an elastic
deformation stage where the conductance stays nearly constant, and
(2) a plastic deformation stage where the configuration is
suddenly changed to one having a smaller cross section, causing
sudden drops of the conductance.

Jumps in the stiffness mostly coincide with the conductance drops,
showing that the jumps are due to the plastic deformation
occurring at the atomic rearrangements. Over the elastic stage,
the stiffness also decreases smoothly on stretching. This shows
that the contact response over the elastic deformation stage is
not linear with the applied stress. The smooth decrease of the
stiffness over a conductance plateau is attributed to bond
weakening upon stretching, as reported for the elongation process
in Au atomic-sized contacts \cite{rubio04}. At the displacement of
1.55nm the stiffness shows a significant change without a clear
conductance drop accompanying it. In fact, the conductance
slightly decreases from 3.4G$_0$ to 3.2G$_0$. Probably this
results from an atomic structural relaxation in the leads near the
central part of the contact that does not affect the contact cross
section, explaining why it is seen in the conductance only as a
minor feature. Similar behavior is also observed around 0.7 nm.

\subsection{Monatomic contact}
\begin{figure}[!t]
\includegraphics[width=7cm]{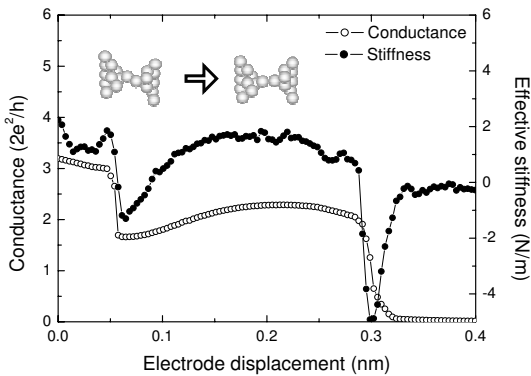}
\caption{The stiffness and conductance measured simultaneously
during stretching of a monatomic Pt contact. The inset shows
schematically the change in configuration from zig-zag to linear
as proposed by Ref.~\onlinecite{garcia05}.}
\label{platinum_two-atom}
\end{figure}
In this section, we will focus on analyzing the mechanical
properties of monatomic contacts. Figure \ref{platinum_two-atom}
shows simultaneous measurement of the stiffness and conductance of
a contact over the last conductance plateau, before breaking. The
conductance decreases to 1.7 $G_0$, which is close to the value
expected for an atomic contact of Pt corresponding to a single
Pt--Pt bond, i.e. a chain of two atoms \cite{nielsen03}. As the
monatomic contact is stretched further the conductance smoothly
increases, then decreases and finally drops to zero when the
contact is broken. Recently, Garc{\'\i}a-Su\'{a}rez \textit{et
al.} studied the electrical transport properties of Pt monatomic
chains suspended by two Pt electrodes using first-principles
simulations \cite{garcia05}. They found that a zigzag arrangement
is the most stable configuration in Pt monatomic chains. According
to Ref.~\onlinecite{garcia05}, the conductance is expected to
increase from 1.5 $G_0$ to 2 $G_0$ when the zigzag arrangement is
stretched into a linear configuration. The conductance increase
from 1.7 $G_0$ to 2.2 $G_0$ in Fig.\ref{platinum_two-atom} might
be attributed to this alignment. When further stretching the
contact the conductance slightly decreases just before breaking,
which is likely due to stretching of the  orbitals
\cite{cuevas98}. Similar to the behavior observed from the
conductance the stiffness initially increases at the last
conductance plateau and then decreases with increasing elongation.
The stiffness of the zigzag arrangement can be expected to be
smaller than that of a linear chain. The final decrease of the
stiffness just before breaking would then result from bond
weakening under extreme strain \cite{rubio04}.

The absolute values for the stiffness measured for Pt atomic
contacts are typically smaller than those reported in molecular
dynamics simulations \cite{pauly06} and also smaller than the
typical experimental values reported for Au atomic contacts
\cite{rubio96, rubio01, rubio04}. Note that the measured value
$\Delta k$ is an effective stiffness which results from the series
connection of the chain itself, the supporting electrodes, and the
interfaces between the chain and electrodes. The stiffness of the
electrodes and the interfaces contributes to the measured value
\cite{pauly06, rubio01} and will be different in each experimental
run. In addition, the two Pt tips on each side of the junction may
be misaligned, touching side-ways. These two effects, elasticity
of the electrodes and misalignment, give rise to a significant
lowering of the effective stiffness measured. Yet, since we find a
clear correlation between the conductance and stiffness behavior,
which is consistent with the results reported previously
\cite{valkering05, rubio04}, the measured variations in stiffness
must reflect the qualitative behavior of the mechanical properties
of the Pt atomic contact correctly.  The absolute value of the
stiffness of the monatomic chain cannot be determined with
precision, and further study will be required on this point.

\subsection{Monatomic chains}
\begin{figure}[!t]
\includegraphics[width=7cm]{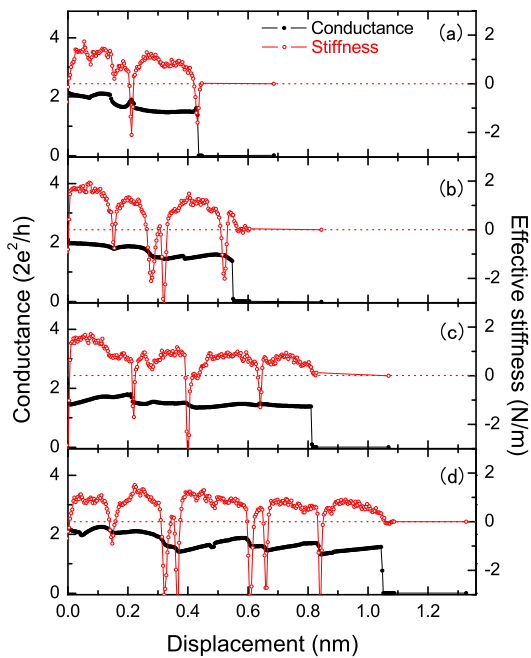}
\caption{Simultaneous measurements of the stiffness and
conductance during the breaking of monatomic Pt chains
corresponding to a chain length of (a) 3, (b) 4, (c) 5 and (d) 6
atoms.} \label{platinum_chain}
\end{figure}
In many cases, by further stretching a monatomic contact, a chain
of atoms can be formed. In Figs.~\ref{platinum_chain}(a)--(d), all
the last conductance plateaus are (much) longer than in the case
of Fig.~\ref{platinum_two-atom}. The conductance is about 2~$G_0$
at the starting point of the plateau, and then shows some
variations until breaking of the contact at about 1.4~$G_0$. The
conductance values on the plateau indicate that the cross section
of the contact is just one atom. The plateaus are stretched over
0.4 nm, 0.6 nm, 0.8 nm and 1.0 nm, corresponding to approximately
three, four, five and six inter-atomic Pt--Pt distances in a
linear arrangement, respectively. These are a strong indication
that monatomic chains have formed between the electrodes, having
one atom in cross-section and up to seven atoms in length. In the
stiffness curve, there are several sudden drops occurring
approximately periodically along the plateau, and the period is
close to the inter-atomic distance of Pt atoms in a chain. The
sudden drops are expected to result form the relaxation of the
force during an atomic rearrangement of the contact, when a new
atom accommodates into the chain. A similar behavior of the
stiffness has been reported for Au monatomic chains by
Rubio-Bollinger $\textit{et al.}$ \cite{rubio04}.

\begin{figure}
\includegraphics[width=8cm]{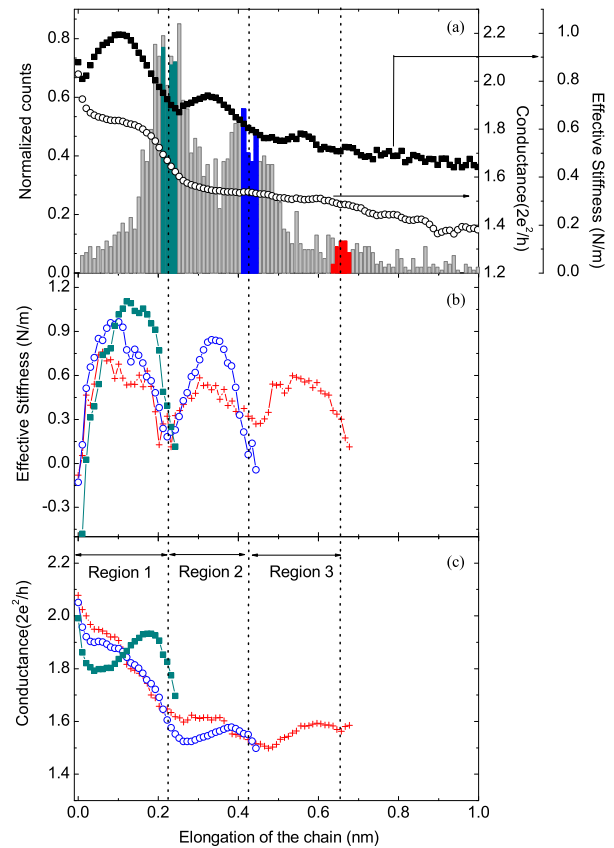}
\caption{(Color online)(a) Length-histogram (left scale, gray
bars) shown together with the average conductance (circles) and
the average stiffness (filled squares) as a function of the length
of Pt monatomic chains. The values obtained for stiffness and
conductance of the chains at each fixed length were averaged over
2200 runs. See text for further details. We consider the colored
areas in (a) to select only those chains that break in a narrow
length range for obtaining the averages shown in (b) for the
stiffness and (c) for the conductance. The three length regimes
range from 0.20 to 0.24 nm (blue), 0.40 to 0.44 nm (green), and
0.63 to 0.67 nm (red).} \label{each_chain}
\end{figure}

In previous studies chain formation has been demonstrated by
analyzing length-histograms, which are obtained by collecting the
distribution of the lengths of the last conductance plateaus
\cite{yanson98,smit01,smit03}. We measured a similar length
histogram from 2200 individual traces, constructed by taking a
conductance dropping below 2.2 $G_0$ as the starting point of the
chain. In addition we measured the average stiffness and
conductance during the breaking of the Pt chains. The length scale
was divided into a number of small bins, and the stiffness and
conductance were averaged in each bin at the given length of the
chains. The average stiffness and conductance as a function of the
chain length are shown in Fig. ~\ref{each_chain} (a). In this
figure, the length-histogram is superimposed. There are two
distinct peaks in the length-histogram, and their central
positions are equally spaced with a spacing of about one
inter-atomic distance, 0.21nm. Looking at the data we first
observe that the average stiffness shows local minima at the peak
positions in the length-histogram. This supports the
interpretation of the length histogram as will be discussed in
more detail below. Note that the conductance does not show the
same period and has an oscillation with double period that was
previously reported and attributed to conductance oscillation as a
function of the even-odd number of atoms in the chain
\cite{smit03}.

Further we have analyzed the behavior of the stiffness as a
function of the number of atoms in the chains. We have divided the
length-histogram into three main regimes, each of which is
centered on the peaks in the length-histogram in order to obtain a
more precise estimate of the number of atoms in the
chain.\cite{footnote} The stiffness and conductance have been
averaged just for those chains breaking at these particular
lengths. The results are shown as a function of the elongation in
Figs.~\ref{each_chain} (b) and (c) for the average stiffness and
conductance, respectively. For region 1, the lengths of the chains
have been selected to be between 0.20 and 0.24 nm, and is
attributed to a two-atom chain. The regions 2 and 3 represent
three- and four-atom chains, defined by lengths ranging from 0.40
to 0.44, and from 0.63 to 0.67, respectively. In Fig. 5 (b) and
(c), the averaged stiffness and conductance for two-atom chains
(filled squares), three-atom chains (circles) and four-atom chains
(crosses) were constructed from 290, 181 and 40 curves,
respectively. When one compares the stiffness as a function of the
number of atoms in the chain the following can be observed:
firstly, the stiffness drops just before breaking in each curve.
This is likely due to bond-weakening, showing that Pt monatomic
chains break from a stretched state rather than a relaxed state.
Secondly, the oscillation period of the stiffness curve is
approximately the inter-atomic distance for all three regimes.
During the process leading to the formation of a three-atom chain
the stiffness shows a local minimum at the first peak in the
length histogram. In the process of formation of a four-atom chain
the stiffness also shows local minima at the first and second
peaks in the length histogram. The positions of the local minima
in stiffness do not depend on the final length of the chain, and
the periodicity of the local minima almost agrees with the Pt--Pt
inter-atomic distance in a linear arrangement. Therefore, the
local minima in stiffness reflect a relaxation of the force acting
on the contact by accommodation of a new atom into the chain. This
supports the idea that the peaks in the length-histogram are a
measure of the number of atoms in the chains. Thirdly, the
stiffness just before breaking decreases with the chain length. To
avoid the effect of the strong drop in stiffness due to the bond
weakening the local maximum of the stiffness just before the bond
weakening can be compared. The local maxima of the stiffness are
about 1.1 N/m for the two-atom chain, 0.8 N/m for three-atom
chain, and 0.6 N/m for four-atom chain. This decrease in stiffness
with length is in qualitative agreement with the result on Au
monatomic chains shown in Ref.~\onlinecite{rubio01}. Fourth, the
position and amplitude of the maximum in the stiffness in a given
length regime depend on the final chain length. Namely, the
smaller the maximum stiffness in a certain regime the longer the
chain can be stretched. This can be seen for all the three sectors
as follows: for the first length regime, the chains that end up
having a maximum length of two atoms have a higher maximum
stiffness than those that end up having a maximum of three or four
atoms and the maximum is found at a longer stretch-length. The
same is valid in the second length regime where the three-atom
chains have a higher stiffness than those ending in four atoms.
One possible explanation for this fact is that a chain can sustain
a maximum force that is independent of its length, as was reported
by force measurements on Au chains \cite{rubio01}. Roughly
speaking, one might integrate the measured stiffness over the
elongation to obtain the total force and, therefore, for the same
elongation a larger force acts on the chain that breaks early, for
which the stiffness grows longer and higher, resulting in breaking
at a shorter length. This is reasonable in view of the mechanism
expected to be involved in the growing of the chain. If the atomic
structure of the leads near the chain section permits releasing
new atoms for addition to the chain length the structure will have
a lower stiffness. Only when the leads are more rigidly ordered
such that no atoms are available for lengthening of the chain can
the force (and the stiffness) in the chain develop its larger
intrinsic value. Finally, in the regions 1 and 2, both the
stiffness and the conductance show an increase followed by a
decrease before the breaking point of the chain. This is
consistent with what is observed in the individual trace of
Fig.~\ref{platinum_two-atom}, and might be attributed to a zigzag
to linear transition.

\section{Conclusion}
The mechanical properties of Pt monatomic chains were studied
using our newly developed MCBJ system with a tuning fork as a
force sensor. The system enables us to measure simultaneously an
effective stiffness and the conductance of the chain, and to
obtain enough data for a statistical analysis. In the shortest
contacts having a single Pt--Pt bond (the two-atom chain) the
stiffness and conductance increases with stretching, which can be
attributed to the transition from zigzag to linear arrangement,
and then decreases due to bond weakening under extreme strain.
From our statistical analysis on longer monatomic chains, it was
found that the stiffness decreases with elongation of the chain,
having local minima due to the bond weakening and subsequent
atomic rearrangements in the chain. Moreover, the minima in the
stiffness appeared at the same positions as the peaks in the
length-histogram. This agrees with the interpretation that the
peaks in the length-histogram are a measure of the number of atoms
in the chain. The final length of the chain depends on its
stiffness, that is, softer chains can grow longer.

\begin{acknowledgments}
This work is part of the research program of the ``Stichting
FOM,'' partly sponsored through the SONS Programme of the European
Science Foundation, which is also funded by the European
Commission, Sixth Framework Programme.
\end{acknowledgments}


\begin{thebibliography}{26}
\expandafter\ifx\csname natexlab\endcsname\relax\def\natexlab#1{#1}\fi
\expandafter\ifx\csname bibnamefont\endcsname\relax
  \def\bibnamefont#1{#1}\fi
\expandafter\ifx\csname bibfnamefont\endcsname\relax
  \def\bibfnamefont#1{#1}\fi
\expandafter\ifx\csname citenamefont\endcsname\relax
  \def\citenamefont#1{#1}\fi
\expandafter\ifx\csname url\endcsname\relax
  \def\url#1{\texttt{#1}}\fi
\expandafter\ifx\csname urlprefix\endcsname\relax\def\urlprefix{URL }\fi
\providecommand{\bibinfo}[2]{#2}
\providecommand{\eprint}[2][]{\url{#2}}

\bibitem[{\citenamefont{Yanson et~al.}(1998)\citenamefont{Yanson, {Rubio
  Bollinger}, van~den Brom, Agra{\"i}t, and van Ruitenbeek}}]{yanson98}
\bibinfo{author}{\bibfnamefont{A.~I.} \bibnamefont{Yanson}},
  \bibinfo{author}{\bibfnamefont{G.}~\bibnamefont{{Rubio Bollinger}}},
  \bibinfo{author}{\bibfnamefont{H.~E.} \bibnamefont{van~den Brom}},
  \bibinfo{author}{\bibfnamefont{N.}~\bibnamefont{Agra{\"i}t}},
  \bibnamefont{and} \bibinfo{author}{\bibfnamefont{J.~M.} \bibnamefont{van
  Ruitenbeek}}, \bibinfo{journal}{Nature} \textbf{\bibinfo{volume}{395}},
  \bibinfo{pages}{783} (\bibinfo{year}{1998}).

\bibitem[{\citenamefont{Ohnishi et~al.}(1998)\citenamefont{Ohnishi, Kondo, and
  Takayanagi}}]{ohnishi98}
\bibinfo{author}{\bibfnamefont{H.}~\bibnamefont{Ohnishi}},
  \bibinfo{author}{\bibfnamefont{Y.}~\bibnamefont{Kondo}}, \bibnamefont{and}
  \bibinfo{author}{\bibfnamefont{K.}~\bibnamefont{Takayanagi}},
  \bibinfo{journal}{Nature} \textbf{\bibinfo{volume}{395}},
  \bibinfo{pages}{780} (\bibinfo{year}{1998}).

\bibitem[{\citenamefont{Agra{\"i}t et~al.}(2003)\citenamefont{Agra{\"i}t, {Levy
  Yeyati}, and van Ruitenbeek}}]{agrait03}
\bibinfo{author}{\bibfnamefont{N.}~\bibnamefont{Agra{\"i}t}},
  \bibinfo{author}{\bibfnamefont{A.}~\bibnamefont{{Levy Yeyati}}},
  \bibnamefont{and} \bibinfo{author}{\bibfnamefont{J.~M.} \bibnamefont{van
  Ruitenbeek}}, \bibinfo{journal}{Phys. Rep.} \textbf{\bibinfo{volume}{377}},
  \bibinfo{pages}{81} (\bibinfo{year}{2003}).

\bibitem[{\citenamefont{Agra{\"\i}t et~al.}(2002)\citenamefont{Agra{\"\i}t,
  Untiedt, Rubio-Bollinger, and Vieira}}]{agrait02}
\bibinfo{author}{\bibfnamefont{N.}~\bibnamefont{Agra{\"\i}t}},
  \bibinfo{author}{\bibfnamefont{C.}~\bibnamefont{Untiedt}},
  \bibinfo{author}{\bibfnamefont{G.}~\bibnamefont{Rubio-Bollinger}},
  \bibnamefont{and} \bibinfo{author}{\bibfnamefont{S.}~\bibnamefont{Vieira}},
  \bibinfo{journal}{Chem. Phys.} \textbf{\bibinfo{volume}{281}},
  \bibinfo{pages}{231} (\bibinfo{year}{2002}).

\bibitem[{\citenamefont{Smit et~al.}(2003)\citenamefont{Smit, Untiedt,
  Rubio-Bollinger, Segers, and van Ruitenbeek}}]{smit03}
\bibinfo{author}{\bibfnamefont{R.~H.~M.} \bibnamefont{Smit}},
  \bibinfo{author}{\bibfnamefont{C.}~\bibnamefont{Untiedt}},
  \bibinfo{author}{\bibfnamefont{G.}~\bibnamefont{Rubio-Bollinger}},
  \bibinfo{author}{\bibfnamefont{R.~C.} \bibnamefont{Segers}},
  \bibnamefont{and} \bibinfo{author}{\bibfnamefont{J.~M.} \bibnamefont{van
  Ruitenbeek}}, \bibinfo{journal}{Phys. Rev. Lett.}
  \textbf{\bibinfo{volume}{91}}, \bibinfo{pages}{076805}
  (\bibinfo{year}{2003}).

\bibitem[{\citenamefont{Stafford}(1998)}]{stafford98}
\bibinfo{author}{\bibfnamefont{C.~A.} \bibnamefont{Stafford}},
  \bibinfo{journal}{Physica E} \textbf{\bibinfo{volume}{1}},
  \bibinfo{pages}{310} (\bibinfo{year}{1998}).

\bibitem[{\citenamefont{Stafford et~al.}(1999)\citenamefont{Stafford, Kassubek,
  B{\"u}rki, and Grabert}}]{stafford99}
\bibinfo{author}{\bibfnamefont{C.~A.} \bibnamefont{Stafford}},
  \bibinfo{author}{\bibfnamefont{F.}~\bibnamefont{Kassubek}},
  \bibinfo{author}{\bibfnamefont{J.}~\bibnamefont{B{\"u}rki}},
  \bibnamefont{and} \bibinfo{author}{\bibfnamefont{H.}~\bibnamefont{Grabert}},
  \bibinfo{journal}{Phys. Rev. Lett.} \textbf{\bibinfo{volume}{83}},
  \bibinfo{pages}{4836} (\bibinfo{year}{1999}).

\bibitem[{\citenamefont{Rubio et~al.}(1996)\citenamefont{Rubio, {Agra{\"i}t},
  and Vieira}}]{rubio96}
\bibinfo{author}{\bibfnamefont{G.}~\bibnamefont{Rubio}},
  \bibinfo{author}{\bibfnamefont{N.}~\bibnamefont{{Agra{\"i}t}}},
  \bibnamefont{and} \bibinfo{author}{\bibfnamefont{S.}~\bibnamefont{Vieira}},
  \bibinfo{journal}{Phys. Rev. Lett.} \textbf{\bibinfo{volume}{76}},
  \bibinfo{pages}{2302} (\bibinfo{year}{1996}).

\bibitem[{\citenamefont{{Rubio-Bollinger}
  et~al.}(2001)\citenamefont{{Rubio-Bollinger}, Bahn, Agra{\"i}t, Jacobsen, and
  Vieira}}]{rubio01}
\bibinfo{author}{\bibfnamefont{G.}~\bibnamefont{{Rubio-Bollinger}}},
  \bibinfo{author}{\bibfnamefont{S.~R.} \bibnamefont{Bahn}},
  \bibinfo{author}{\bibfnamefont{N.}~\bibnamefont{Agra{\"i}t}},
  \bibinfo{author}{\bibfnamefont{K.~W.} \bibnamefont{Jacobsen}},
  \bibnamefont{and} \bibinfo{author}{\bibfnamefont{S.}~\bibnamefont{Vieira}},
  \bibinfo{journal}{Phys. Rev. Lett.} \textbf{\bibinfo{volume}{87}},
  \bibinfo{pages}{026101} (\bibinfo{year}{2001}).

\bibitem[{\citenamefont{Valkering et~al.}(2005)\citenamefont{Valkering, Mares,
  Untiedt, {Babaei Gavan}, Oosterkamp, and van Ruitenbeek}}]{valkering05}
\bibinfo{author}{\bibfnamefont{A.~M.~C.} \bibnamefont{Valkering}},
  \bibinfo{author}{\bibfnamefont{A.~I.} \bibnamefont{Mares}},
  \bibinfo{author}{\bibfnamefont{C.}~\bibnamefont{Untiedt}},
  \bibinfo{author}{\bibfnamefont{K.}~\bibnamefont{{Babaei Gavan}}},
  \bibinfo{author}{\bibfnamefont{T.~H.} \bibnamefont{Oosterkamp}},
  \bibnamefont{and} \bibinfo{author}{\bibfnamefont{J.~M.} \bibnamefont{van
  Ruitenbeek}}, \bibinfo{journal}{Rev. Sci. Instrum.}
  \textbf{\bibinfo{volume}{76}}, \bibinfo{pages}{103903}
  (\bibinfo{year}{2005}).

\bibitem[{\citenamefont{{Rubio-Bollinger}
  et~al.}(2004)\citenamefont{{Rubio-Bollinger}, Joyez, and
  Agra{\"i}t}}]{rubio04}
\bibinfo{author}{\bibfnamefont{G.}~\bibnamefont{{Rubio-Bollinger}}},
  \bibinfo{author}{\bibfnamefont{P.}~\bibnamefont{Joyez}}, \bibnamefont{and}
  \bibinfo{author}{\bibfnamefont{N.}~\bibnamefont{Agra{\"i}t}},
  \bibinfo{journal}{Phys. Rev. Lett.} \textbf{\bibinfo{volume}{93}},
  \bibinfo{pages}{116803} (\bibinfo{year}{2004}).

\bibitem[{\citenamefont{Torres et~al.}(1999)\citenamefont{Torres, Tosatti,
  Corso, Ercolessi, Kohanoff, Tolla, and Soler}}]{Torres99}
\bibinfo{author}{\bibfnamefont{J.~A.} \bibnamefont{Torres}},
  \bibinfo{author}{\bibfnamefont{E.}~\bibnamefont{Tosatti}},
  \bibinfo{author}{\bibfnamefont{A.~D.} \bibnamefont{Corso}},
  \bibinfo{author}{\bibfnamefont{F.}~\bibnamefont{Ercolessi}},
  \bibinfo{author}{\bibfnamefont{J.~J.} \bibnamefont{Kohanoff}},
  \bibinfo{author}{\bibfnamefont{F.~D.~D.} \bibnamefont{Tolla}},
  \bibnamefont{and} \bibinfo{author}{\bibfnamefont{J.~M.} \bibnamefont{Soler}},
  \bibinfo{journal}{Surface Science} \textbf{\bibinfo{volume}{426}},
  \bibinfo{pages}{L441} (\bibinfo{year}{1999}).

\bibitem[{\citenamefont{Todorov and Sutton}(1996)}]{Todorov96}
\bibinfo{author}{\bibfnamefont{T.~N.} \bibnamefont{Todorov}} \bibnamefont{and}
  \bibinfo{author}{\bibfnamefont{A.~P.} \bibnamefont{Sutton}},
  \bibinfo{journal}{Phys. Rev. B} \textbf{\bibinfo{volume}{54}},
  \bibinfo{pages}{R14234} (\bibinfo{year}{1996}).

\bibitem[{\citenamefont{Sorensen et~al.}(1998)\citenamefont{Sorensen,
  Brandbyge, and Jacobsen}}]{sorensen98}
\bibinfo{author}{\bibfnamefont{M.~R.} \bibnamefont{Sorensen}},
  \bibinfo{author}{\bibfnamefont{M.}~\bibnamefont{Brandbyge}},
  \bibnamefont{and} \bibinfo{author}{\bibfnamefont{K.~W.}
  \bibnamefont{Jacobsen}}, \bibinfo{journal}{Phys. Rev. B}
  \textbf{\bibinfo{volume}{57}}, \bibinfo{pages}{3283} (\bibinfo{year}{1998}).

\bibitem[{\citenamefont{S{\'a}nchez-Portal
  et~al.}(1999)\citenamefont{S{\'a}nchez-Portal, Artacho, Junquera,
  Ordej{\'o}n, Garc{\'i}a, and Soler}}]{sanchez99}
\bibinfo{author}{\bibfnamefont{D.}~\bibnamefont{S{\'a}nchez-Portal}},
  \bibinfo{author}{\bibfnamefont{E.}~\bibnamefont{Artacho}},
  \bibinfo{author}{\bibfnamefont{J.}~\bibnamefont{Junquera}},
  \bibinfo{author}{\bibfnamefont{P.}~\bibnamefont{Ordej{\'o}n}},
  \bibinfo{author}{\bibfnamefont{A.}~\bibnamefont{Garc{\'i}a}},
  \bibnamefont{and} \bibinfo{author}{\bibfnamefont{J.~M.} \bibnamefont{Soler}},
  \bibinfo{journal}{Phys. Rev. Lett.} \textbf{\bibinfo{volume}{83}},
  \bibinfo{pages}{3884} (\bibinfo{year}{1999}).

\bibitem[{\citenamefont{H{\"a}kkinen et~al.}(2000)\citenamefont{H{\"a}kkinen,
  Barnett, Scherbakov, and Landman}}]{hakkinen00}
\bibinfo{author}{\bibfnamefont{H.}~\bibnamefont{H{\"a}kkinen}},
  \bibinfo{author}{\bibfnamefont{R.~N.} \bibnamefont{Barnett}},
  \bibinfo{author}{\bibfnamefont{A.~G.} \bibnamefont{Scherbakov}},
  \bibnamefont{and} \bibinfo{author}{\bibfnamefont{U.}~\bibnamefont{Landman}},
  \bibinfo{journal}{J. Phys. Chem. B} \textbf{\bibinfo{volume}{104}},
  \bibinfo{pages}{9063} (\bibinfo{year}{2000}).

\bibitem[{\citenamefont{da~Silva et~al.}(2001)\citenamefont{da~Silva, da~Silva,
  and Fazzio}}]{Silva01}
\bibinfo{author}{\bibfnamefont{E.~Z.} \bibnamefont{da~Silva}},
  \bibinfo{author}{\bibfnamefont{A.~J.~R.} \bibnamefont{da~Silva}},
  \bibnamefont{and} \bibinfo{author}{\bibfnamefont{A.}~\bibnamefont{Fazzio}},
  \bibinfo{journal}{Phys. Rev. Lett.} \textbf{\bibinfo{volume}{87}},
  \bibinfo{pages}{256102} (\bibinfo{year}{2001}).

\bibitem[{\citenamefont{Dreher et~al.}(2005)\citenamefont{Dreher, Pauly,
  Heurich, Cuevas, Scheer, and Nielaba}}]{Dreher05}
\bibinfo{author}{\bibfnamefont{M.}~\bibnamefont{Dreher}},
  \bibinfo{author}{\bibfnamefont{F.}~\bibnamefont{Pauly}},
  \bibinfo{author}{\bibfnamefont{J.}~\bibnamefont{Heurich}},
  \bibinfo{author}{\bibfnamefont{J.~C.} \bibnamefont{Cuevas}},
  \bibinfo{author}{\bibfnamefont{E.}~\bibnamefont{Scheer}}, \bibnamefont{and}
  \bibinfo{author}{\bibfnamefont{P.}~\bibnamefont{Nielaba}},
  \bibinfo{journal}{Phys. Rev. B} \textbf{\bibinfo{volume}{72}},
  \bibinfo{pages}{075435} (\bibinfo{year}{2005}).

\bibitem[{\citenamefont{Smit et~al.}(2001)\citenamefont{Smit, Untiedt, Yanson,
  and van Ruitenbeek}}]{smit01}
\bibinfo{author}{\bibfnamefont{R.~H.~M.} \bibnamefont{Smit}},
  \bibinfo{author}{\bibfnamefont{C.}~\bibnamefont{Untiedt}},
  \bibinfo{author}{\bibfnamefont{A.~I.} \bibnamefont{Yanson}},
  \bibnamefont{and} \bibinfo{author}{\bibfnamefont{J.~M.} \bibnamefont{van
  Ruitenbeek}}, \bibinfo{journal}{Phys. Rev. Lett.}
  \textbf{\bibinfo{volume}{87}}, \bibinfo{pages}{266102}
  (\bibinfo{year}{2001}).

\bibitem[{\citenamefont{Bahn and Jacobsen}(2001)}]{bahn01}
\bibinfo{author}{\bibfnamefont{S.~R.} \bibnamefont{Bahn}} \bibnamefont{and}
  \bibinfo{author}{\bibfnamefont{K.~W.} \bibnamefont{Jacobsen}},
  \bibinfo{journal}{Phys. Rev. Lett.} \textbf{\bibinfo{volume}{87}},
  \bibinfo{pages}{266101} (\bibinfo{year}{2001}).

\bibitem[{\citenamefont{Garc{\'i}a-Su{\'a}rez
  et~al.}(2005)\citenamefont{Garc{\'i}a-Su{\'a}rez, Rocha, Bailey, Lambert,
  Sanvito, and Ferrer}}]{garcia05}
\bibinfo{author}{\bibfnamefont{V.~M.} \bibnamefont{Garc{\'i}a-Su{\'a}rez}},
  \bibinfo{author}{\bibfnamefont{A.~R.} \bibnamefont{Rocha}},
  \bibinfo{author}{\bibfnamefont{S.~W.} \bibnamefont{Bailey}},
  \bibinfo{author}{\bibfnamefont{C.~J.} \bibnamefont{Lambert}},
  \bibinfo{author}{\bibfnamefont{S.}~\bibnamefont{Sanvito}}, \bibnamefont{and}
  \bibinfo{author}{\bibfnamefont{J.}~\bibnamefont{Ferrer}},
  \bibinfo{journal}{Phys. Rev. Lett.} \textbf{\bibinfo{volume}{95}},
  \bibinfo{pages}{256804} (\bibinfo{year}{2005}).

\bibitem[{\citenamefont{Pauly et~al.}(2006)\citenamefont{Pauly, Dreher, Viljas,
  Hafner, Cuevas, and Nielaba}}]{pauly06}
\bibinfo{author}{\bibfnamefont{F.}~\bibnamefont{Pauly}},
  \bibinfo{author}{\bibfnamefont{M.}~\bibnamefont{Dreher}},
  \bibinfo{author}{\bibfnamefont{J.~K.} \bibnamefont{Viljas}},
  \bibinfo{author}{\bibfnamefont{M.}~\bibnamefont{Hafner}},
  \bibinfo{author}{\bibfnamefont{J.~C.} \bibnamefont{Cuevas}},
  \bibnamefont{and} \bibinfo{author}{\bibfnamefont{P.}~\bibnamefont{Nielaba}},
  \bibinfo{journal}{Phys. Rev. B} \textbf{\bibinfo{volume}{74}},
  \bibinfo{pages}{235106} (\bibinfo{year}{2006}).

\bibitem[{Lab()}]{LabView}
\bibinfo{note}{LabView 7.0, National Instruments, Austin, TX, USA}.

\bibitem[{\citenamefont{Nielsen et~al.}(2003)\citenamefont{Nielsen, Noat,
  Brandbyge, Smit, Hansen, Chen, Yanson, Besenbacher, and van
  Ruitenbeek}}]{nielsen03}
\bibinfo{author}{\bibfnamefont{S.~K.} \bibnamefont{Nielsen}},
  \bibinfo{author}{\bibfnamefont{Y.}~\bibnamefont{Noat}},
  \bibinfo{author}{\bibfnamefont{M.}~\bibnamefont{Brandbyge}},
  \bibinfo{author}{\bibfnamefont{R.~H.~M.} \bibnamefont{Smit}},
  \bibinfo{author}{\bibfnamefont{K.}~\bibnamefont{Hansen}},
  \bibinfo{author}{\bibfnamefont{L.~Y.} \bibnamefont{Chen}},
  \bibinfo{author}{\bibfnamefont{A.~I.} \bibnamefont{Yanson}},
  \bibinfo{author}{\bibfnamefont{F.}~\bibnamefont{Besenbacher}},
  \bibnamefont{and} \bibinfo{author}{\bibfnamefont{J.~M.} \bibnamefont{van
  Ruitenbeek}}, \bibinfo{journal}{Phys. Rev. B} \textbf{\bibinfo{volume}{67}},
  \bibinfo{pages}{245411} (\bibinfo{year}{2003}).

\bibitem[{\citenamefont{Cuevas et~al.}(1998)\citenamefont{Cuevas, {Levy
  Yeyati}, Martin-Rodero, Bollinger, Untiedt, and Agra{\"i}t}}]{cuevas98}
\bibinfo{author}{\bibfnamefont{J.~C.} \bibnamefont{Cuevas}},
  \bibinfo{author}{\bibfnamefont{A.}~\bibnamefont{{Levy Yeyati}}},
  \bibinfo{author}{\bibfnamefont{A.}~\bibnamefont{Martin-Rodero}},
  \bibinfo{author}{\bibfnamefont{G.~R.} \bibnamefont{Bollinger}},
  \bibinfo{author}{\bibfnamefont{C.}~\bibnamefont{Untiedt}}, \bibnamefont{and}
  \bibinfo{author}{\bibfnamefont{N.}~\bibnamefont{Agra{\"i}t}},
  \bibinfo{journal}{Phys. Rev. Lett.} \textbf{\bibinfo{volume}{81}},
  \bibinfo{pages}{2990} (\bibinfo{year}{1998}).

\bibitem[{foo()}]{footnote}
\bibinfo{note}{The third peak is not very well developed in this data set, but
  its position was verified from a much larger set of data collected
  independently of the force measurement run, for the same contact.}

\end{thebibliography}
\end{document}